\documentclass[a4paper]{jpconf}
\usepackage{graphicx}
\usepackage{xcolor}
\usepackage{amsmath}  
\usepackage{amsfonts} 

\begin{document}
\title{Light illuminance does not always decay with the inverse of the square of the distance: an open-inquiry laboratory}

\author{Cecilia Stari$^{0000-0002-4349-1036, 1}$, Marcos Abreu$^1$, Martín Monteiro$^{0000-0001-9472-2116, 2}$,  Arturo C. Mart{\'i}$^{0000-0003-2023-8676, 1}$ }

\address{$^1$ Instituto de F\'{i}sica, Universidad de la
  Rep\'{u}blica, Montevideo, Uruguay}
\address{$^2$ Universidad ORT Uruguay, Montevideo, Uruguay}

\ead{marti@fisica.edu.uy}

\begin{abstract}
The square inverse law with distance plays an important role in many fields of physics covering electromagnetism, optics or acoustics. However, as every law in physics has its range of validity. We propose an open-inquiry laboratory where we challenge these concepts by proposing experiments where the illuminance of light decays linearly or even remains constant over a range of distances. Using the light sensors built into smartphones, it is possible to measure light curves for different sources: point, linear, planar and even LED ring lights.  The analysis of these curves allows us to discuss the limits of the physical theories. This low-cost laboratory, initially proposed in the context of the COVID19 pandemic, has the virtue of challenging intuition and encouraging the critical spirit of the students.
\end{abstract}

\section{Introduction}

The decay of the illuminance of punctual sources with the inverse of the square of the distance is a consequence of fundamental physical principles \cite{born2013principles}. This dependence is observed not only in optics, but also in electromagnetism \cite{yan1995near} or fluids \cite{schneider2004discovery}, and is generally related to other equally important aspects of classical physics.  Students of science and engineering are repeatedly exposed to fields that decay in this way during their education \cite{voudoukis2017inverse}.
However, as with all models, there are limitations, and discussing the limits of validity is key to developing critical and independent students.

One way to explore the limits of these fundamental laws is to study the decay of illuminance from sources of different shapes and extensions.
In recent years, several experiments have been proposed to study these phenomena using modern technology \cite{downie2007data,bohacek2011using,bates2013inverse}. In Ref. \cite{bohacek2011using} it was proposed to use a laptop screen as a light source and a light sensor designed for educational purposes. By covering the screen with a black cardboard, it is possible to cut out an elongated or circular slit to obtain light sources with different shapes. By placing the sensor manually at different distances from the source, it is possible to study the illuminance decay curves. In another proposal \cite{bates2013inverse}, an inclined rail and a distance sensor are used to automate the measurement process by placing the light sensor on a motion cart that slides on the rail.

In the last decade, a large number of papers have been published with proposals for experimental activities that can be carried out using smartphone sensors \cite{monteiro2022resource}.  In particular, the light sensor built into smartphones has also been used to measure light curves from different sources \cite{vieira2014inverso,salinas2018characterization}.
Smartphone sensors are perfect for home-lab activities, allowing students to measure various physical parameters such as light illuminance, acceleration, or sound levels with ease and accuracy
\cite{salinas2020dynamics,monteiro2022home,monteiro2018bottle}. Additionally, these sensors can be used simultaneously, enabling collaborative experiments and comparative analysis from multiple perspectives, enhancing the depth of scientific inquiry \cite{monteiro2019physics}. Furthermore, combining sensor data with video analysis provides a powerful tool for studying motion, dynamics, and other phenomena, offering a comprehensive approach to experimentation \cite{monteiro2021allies}.

In recent years, the COVID19 pandemic has disrupted educational structures and forced us to adapt to health restrictions \cite{o2021guide}. Once these restrictions were overcome, some aspects were adapted to the new normality, but we also learned to adopt more flexible methods and to look for ways to make better use of classroom time. One of the experimental projects we proposed to our students was the study of light decay curves for sources with different geometries.
We suggested a non-exhausted list comprising: punctual sources (or light bulbs), linear sources (or LED strips), flat sources (or computer monitors) and finally, circular sources (or ring lights such as those available for home video recording) or other light sources the students have at their  disposal.
The project also involved proposing a model to explain the experimental results and discussing the limits of validity.

Once completed, this activity received positive feedback from the students and the group of teachers involved. In the next section we briefly review the theoretical background, then present examples based on the students' own experimental results and discuss the proposed models. The last section outlines the conclusions.

\section{Theoretical background}

From everyday experience, we know that what we colloquially call the “brightness” of a light source decreases as we move farther away from it. In photometry, this familiar perception is quantified using the concept of illuminance, which measures the luminous flux incident per unit area and is expressed in lux (lumens per square meter) within the International System of Units.

The decrease in illuminance with distance is often described by the inverse-square law, which states that for an ideal point source, illuminance decreases proportionally to the square of the distance. This behavior arises from fundamental conservation principles as light spreads out in space, but it also depends on several factors, most notably the geometry of the light source.

Ambient light sensors, commonly integrated into smartphones and portable devices, utilize this measurement to detect changes in environmental lighting conditions and automatically adjust screen brightness. While luminous flux is typically weighted according to the eye’s sensitivity across different wavelengths, in this study we focus exclusively on white light sources, allowing us to disregard spectral weighting and simplify the analysis.

For punctual sources emitting light uniformly in all directions, wavefronts form concentric spheres of radius $r$. In this case, illuminance at a distance $r$ equals the luminous flux divided by the area of the sphere ($4 \pi r^2$), causing the illuminance to decrease with the square of the distance. For instance, doubling the distance reduces the illuminance to one-fourth of its original value.

However, this law has limits of validity. When the sensor is very close to the source, the source cannot be approximated as a point. Conversely, if the source is too far, the illuminance may fall below the sensor's detection threshold. For other geometries, the illuminance decay behavior differs significantly:
\begin{itemize}
    \item 
    Linear sources (e.g., LED strips): illuminance decreases inversely with distance.
    \item 
    Flat sources: illuminance remains nearly constant as long as the distance is smaller than the source size.
   \item  Ring-shaped sources: exhibit a unique non-monotonic pattern, with maximum illuminance at an intermediate distance, followed by a rapid decrease.

\end{itemize}
These distinct decay patterns are critical for understanding and modeling the interaction between light sources and sensors in technological applications.

\section{Experimental setup}
This experimental activity allows great versatility in the design of the set up. It is possible to use commercial lamps as light sources or to use other sources, punctual, linear, plane, and circular, which the students have available either in the laboratory or at home. Additionally, a personal computer or television monitor may be employed as a light source, with black cardboard covering the desired area so that the rest defines the shape and size of the light source. This approach enables the dimensions of the source to be precisely controlled, enabling  an investigation into the validity of the model in terms of the relationship between distance and source dimensions.

The Phyphox app is an impressive tool for effortlessly acquiring experimental data using smartphone sensors \cite{staacks_2018}. It enables students to collect, visualize and export data seamlessly, fostering collaboration by allowing easy sharing among peers. The built-in features of the app help perform statistical analyzes, providing students with a deeper understanding of measurement limitations \cite{monteiro2021using}.

\begin{figure}[htb]
\begin{center}
\includegraphics[width=0.26\textwidth]{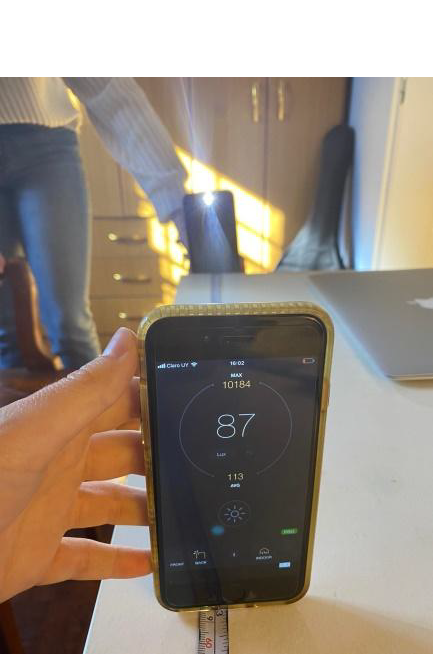}
\includegraphics[width=0.615\textwidth]{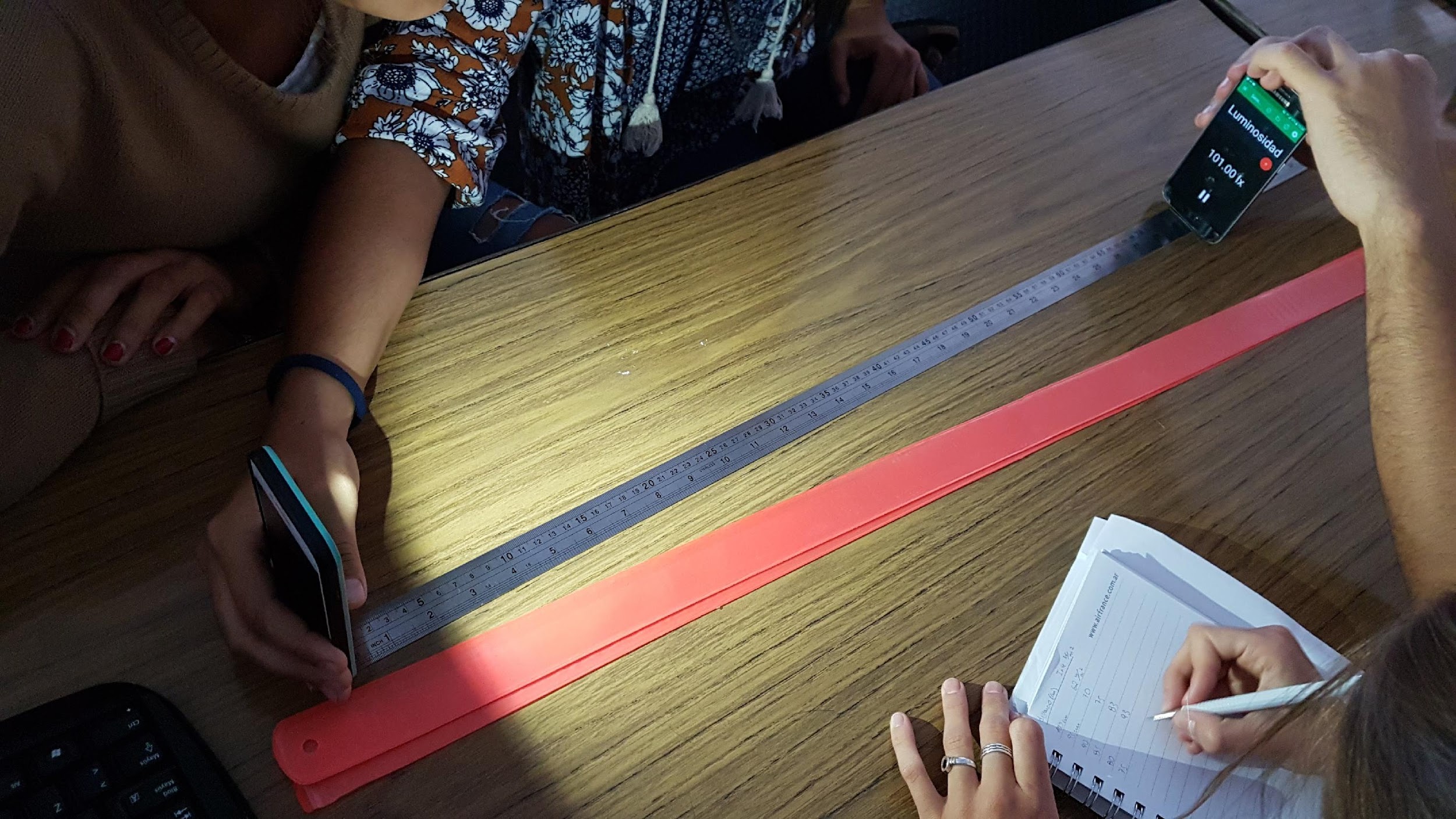}
\caption{Images of our students measuring the light intensity of punctual light sources.
Note that smartphones are used as both source and detector.}
\label{fig:puntual}
\end{center}
\end{figure}

\section{Results obtained by the students}

In this proposal, we consider four light sources. In the first, we study the curves of a punctual source as shown in Fig.~\ref{fig:puntual}, consisting of a light source, a tape measure and a smartphone. The system needs to be aligned correctly to give accurate results. In all cases, illuminance measurements are made using the light sensor of a mobile phone, recording these values as a function of the distance to the source \cite{monteiro2017polarization}. Students can analyze and discuss the validity of the proposed model in each of the situations in relation to the distance to the source and also to the dimensions of the source. For each selected position, several measurements of illuminance were taken over a period of time using Phyphox app, determining the average and standard deviation in real time.

\begin{figure}[h]
\begin{center}
\includegraphics[width=0.95\textwidth]{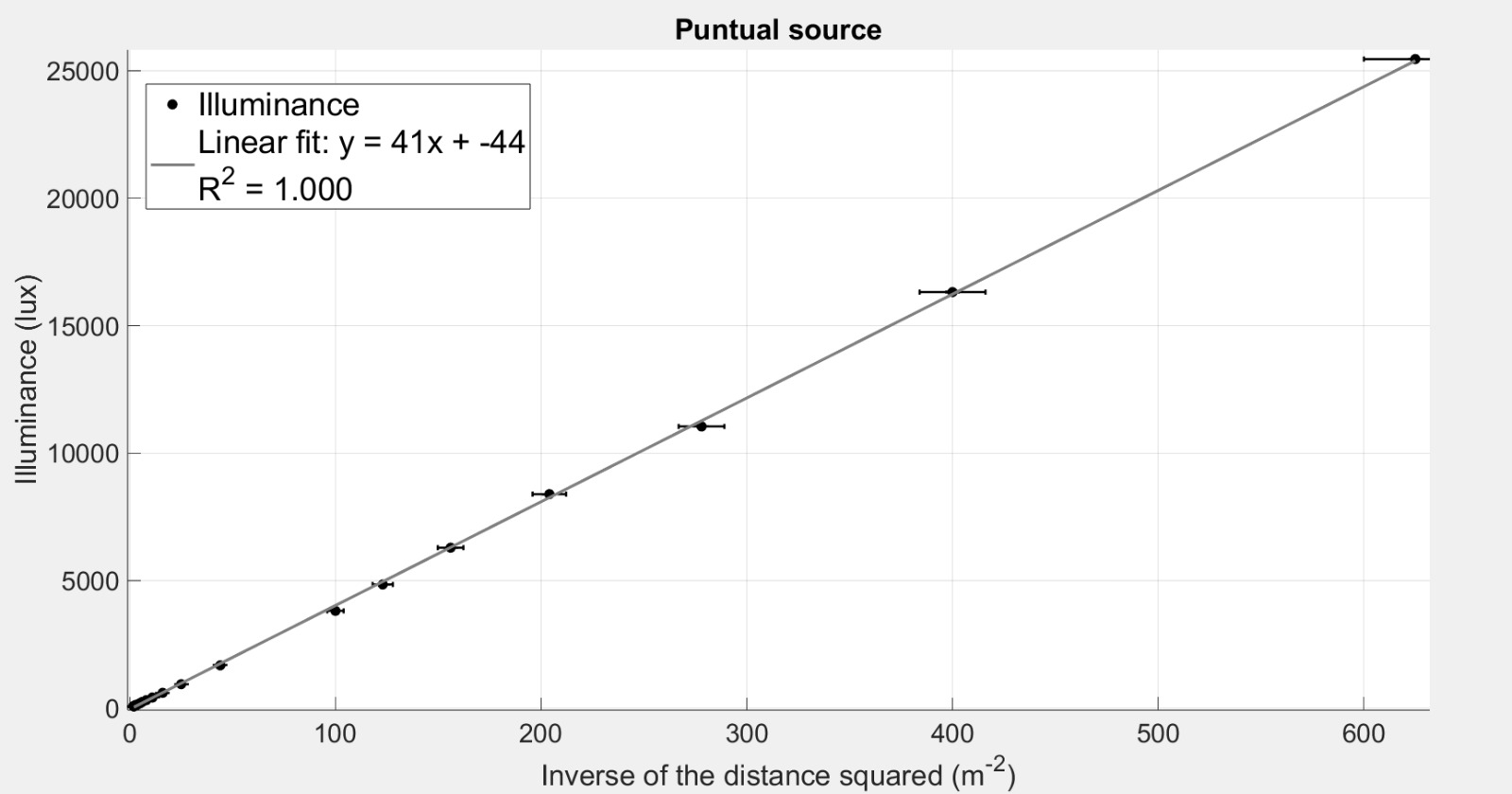}
\caption{Illuminance of point-like source: experimental data and linear fit.}
\label{fig:recta}
\end{center}
\end{figure}

The graphs of illuminance as a function of distance, shown in Fig. \ref{fig:puntual}, allow us to formulate models for each type of source and also to study the validity ranges of each of them. 
These data, and all the results presented here, were obtained by small group of students working at home. The discussions with all group offer opportunity to discuss the models proposed and the range of validity.

\begin{figure}[h]
\begin{center}
\includegraphics[width=0.231\textwidth]{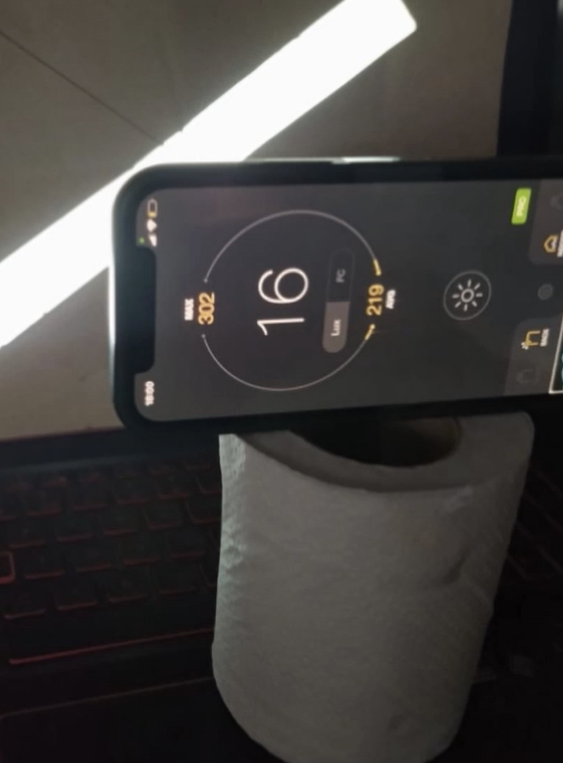}
\includegraphics[width=0.152\textwidth]{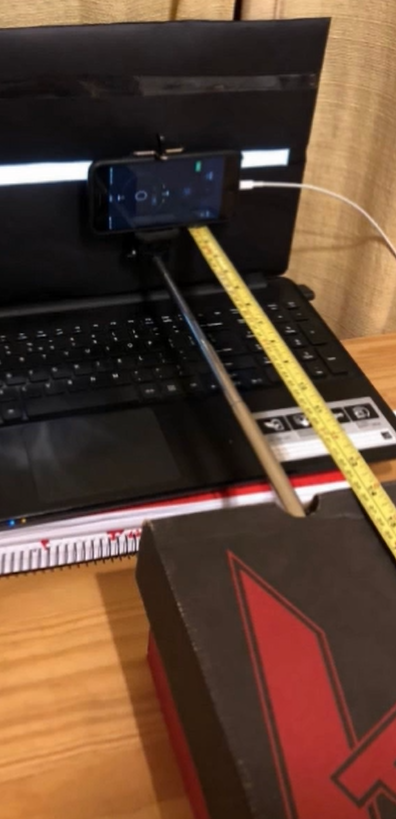}
\includegraphics[width=0.456\textwidth]{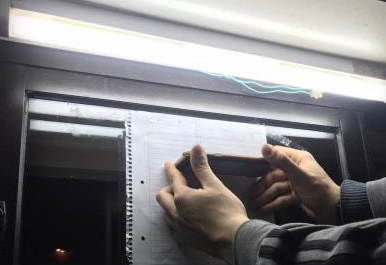}
\caption{Experiment to measure the illuminance of a linear light source.
One alternative is to cover the screen of a monitor showing an all-white slide with black cardboard, except for an elongated slit to form a linear light source.
Another alternative to a linear light source is to use a tube light, which is readily available in shops.}
\label{fig:linear}
\end{center}
\end{figure}

The next situation considered concerns linear light sources, in this case obtained by means of LED strips that are easily accessible to students. An example of the experimental setup is shown in Fig.~\ref{fig:linear}.

\begin{figure}[h]
\begin{center}
\includegraphics[width=0.84\textwidth]{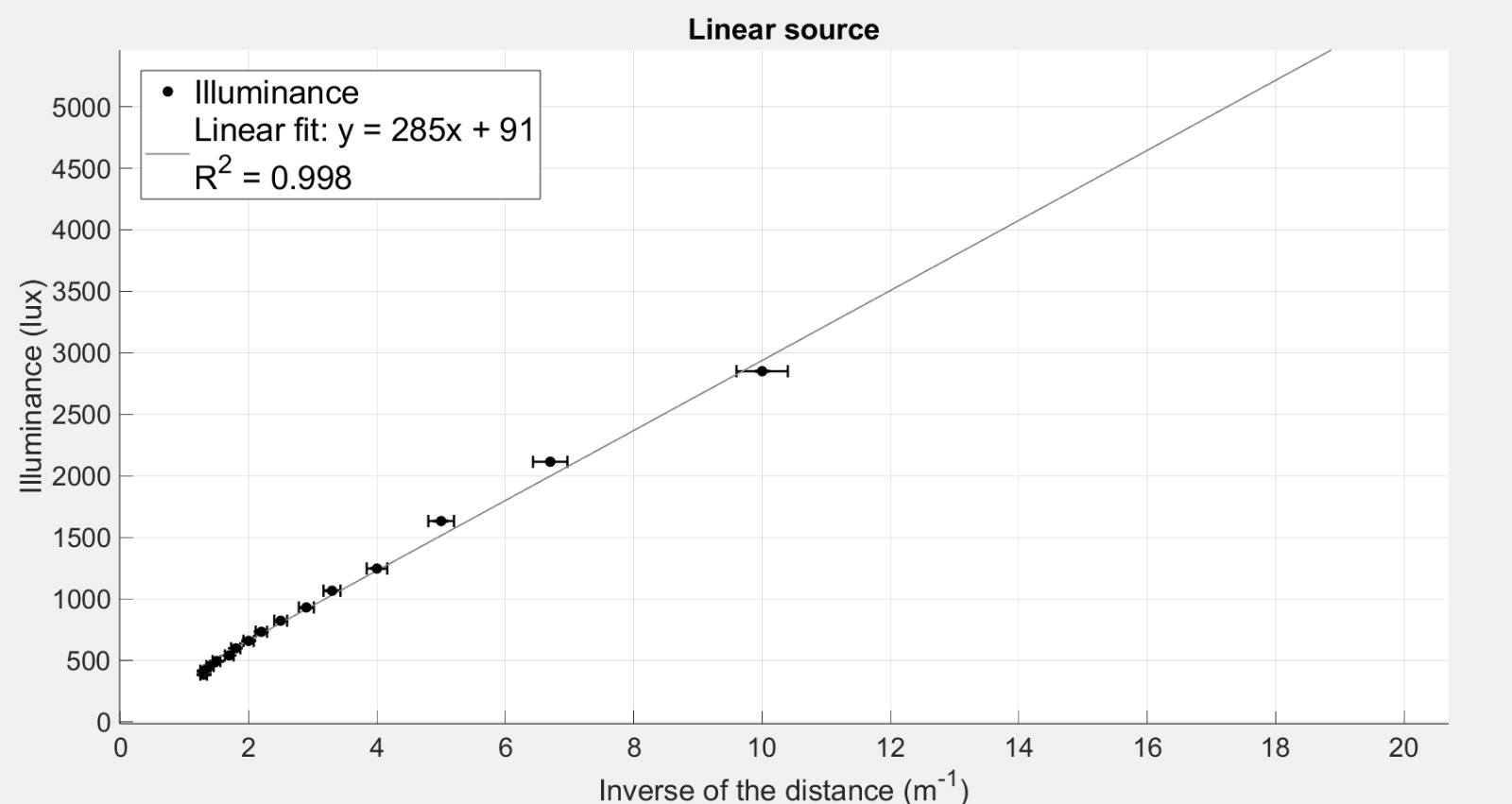}
\caption{Illuminance of a linear source as  function of the inverse of distance. Experimental data are shown with uncertainty bars and linear fit.}
\label{fig:pla3}
\end{center}
\end{figure}

An example of the experimental results is shown in Fig.~\ref{fig:pla3}. In this case, the illuminance is plotted as a function of the inverse of the distance. The linear fit indicates the validity of the model over a reasonable range of distances.

\begin{figure}[h]
\begin{center}
\includegraphics[width=0.39\textwidth]{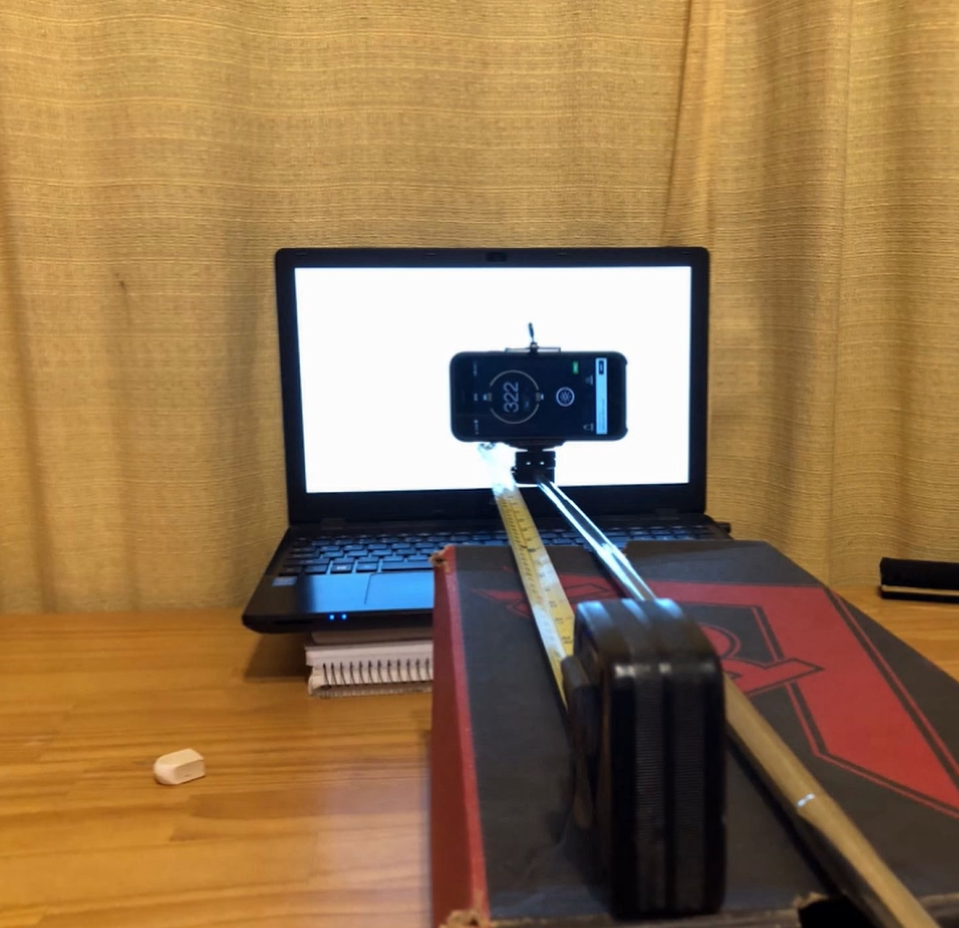}
\includegraphics[width=0.593\textwidth]{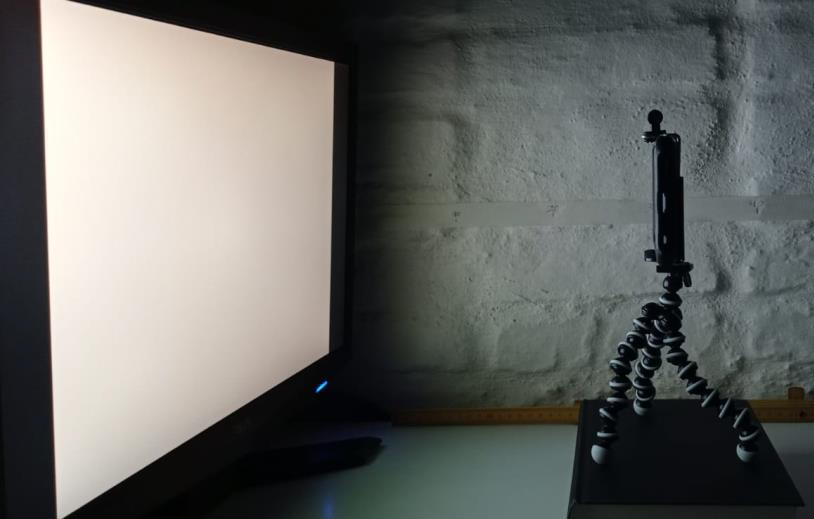}
\caption{With a screen display and a completely blank slide, a planar light source is easily obtained.
Here we show you two experimental set-ups that were effectively used by the students.}
\label{fig:planar}
\end{center}
\end{figure}

Flat screens or televisions were also used by our students. In this case, we consider the screen at maximum brightness with a white slide. The smartphone was placed on a stand at different distances from the source, as shown in Fig.~\ref{fig:planar}. 

\begin{figure}[h]
\begin{center}
\includegraphics[width=0.74\textwidth]{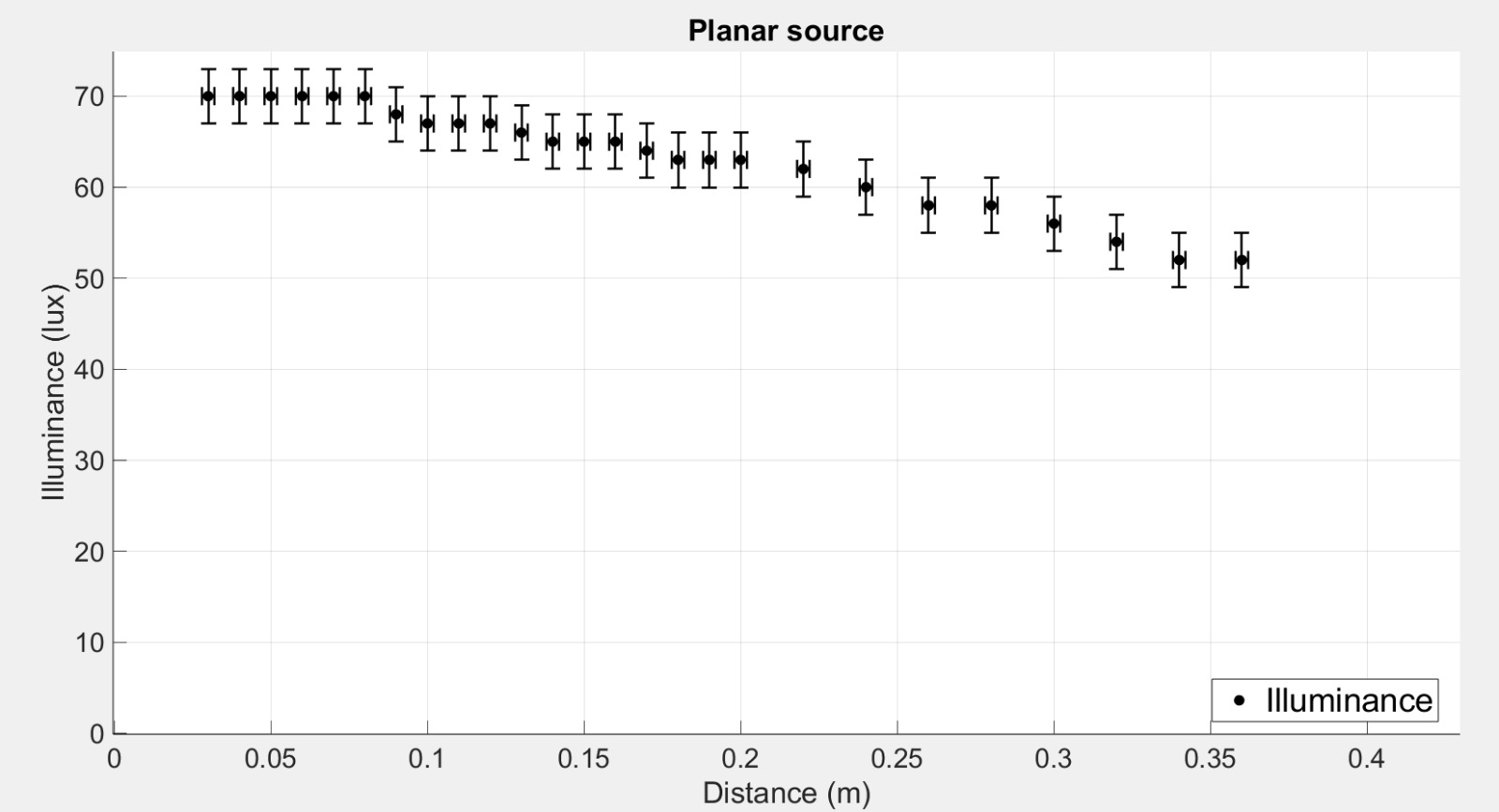}
\caption{Illuminance as a function of distance for a flat source. It can be seen that close to the source the illuminance remains constant before it begins to decay smoothly.}
\label{fig:pla4}
\end{center}
\end{figure}

An example of the results obtained for the flat source is shown in Fig.\ref{fig:pla4}. We observe an initial range of distances where the illuminance remains constant and then starts to slowly decrease after a certain distance. This distance is related to the size of the monitor and corresponds to the limit of validity of the model.

\begin{figure}[h]
\begin{center}
\includegraphics[width=0.74\textwidth]{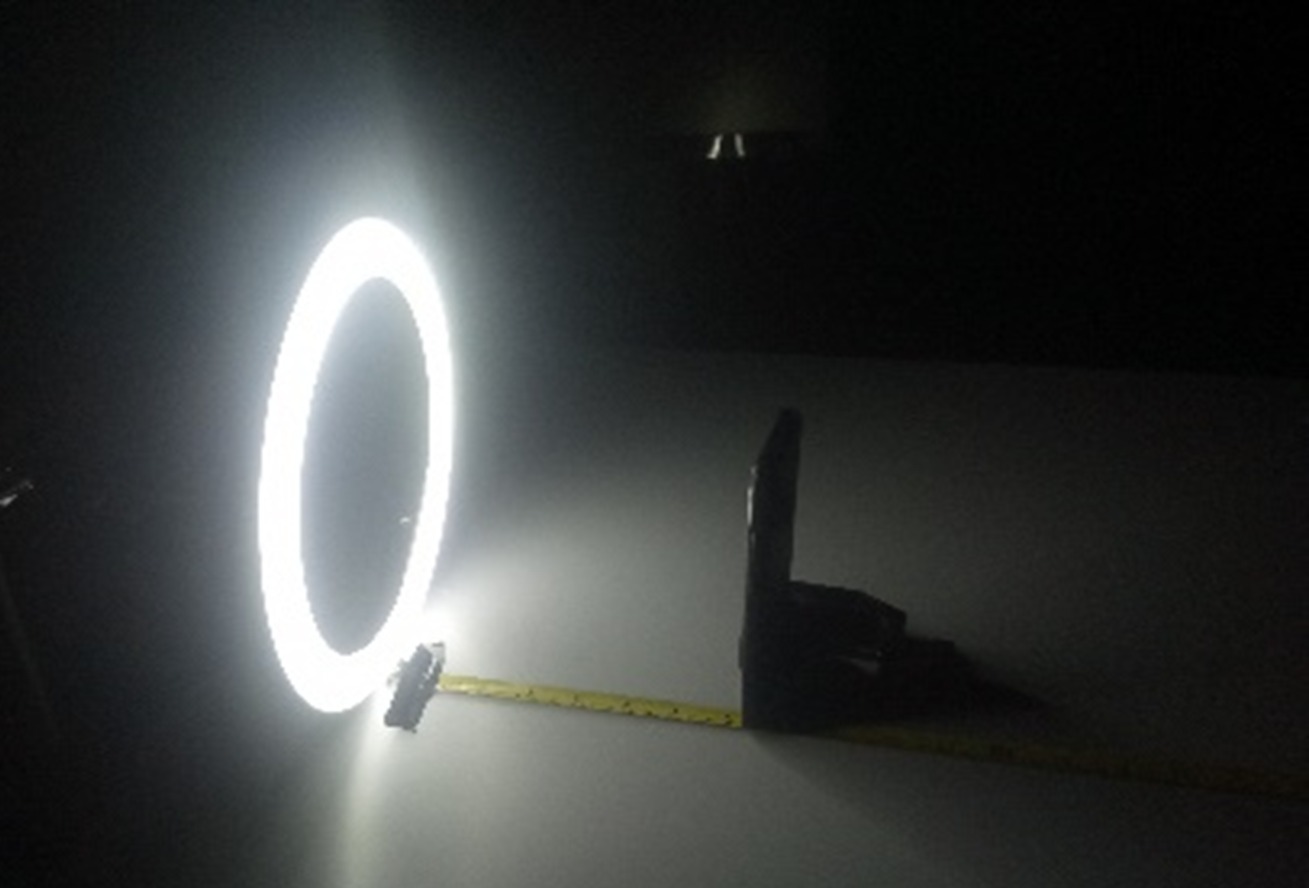}
\caption{A circular light source used in many places for live web presentations
is an unconventional geometry that stimulates student interest by encouraging discussion of
the proposed models and their limitations.}
\label{fig:ringsetup}
\end{center}
\end{figure}

\begin{figure}[h]
\begin{center}
\includegraphics[width=0.74\textwidth]{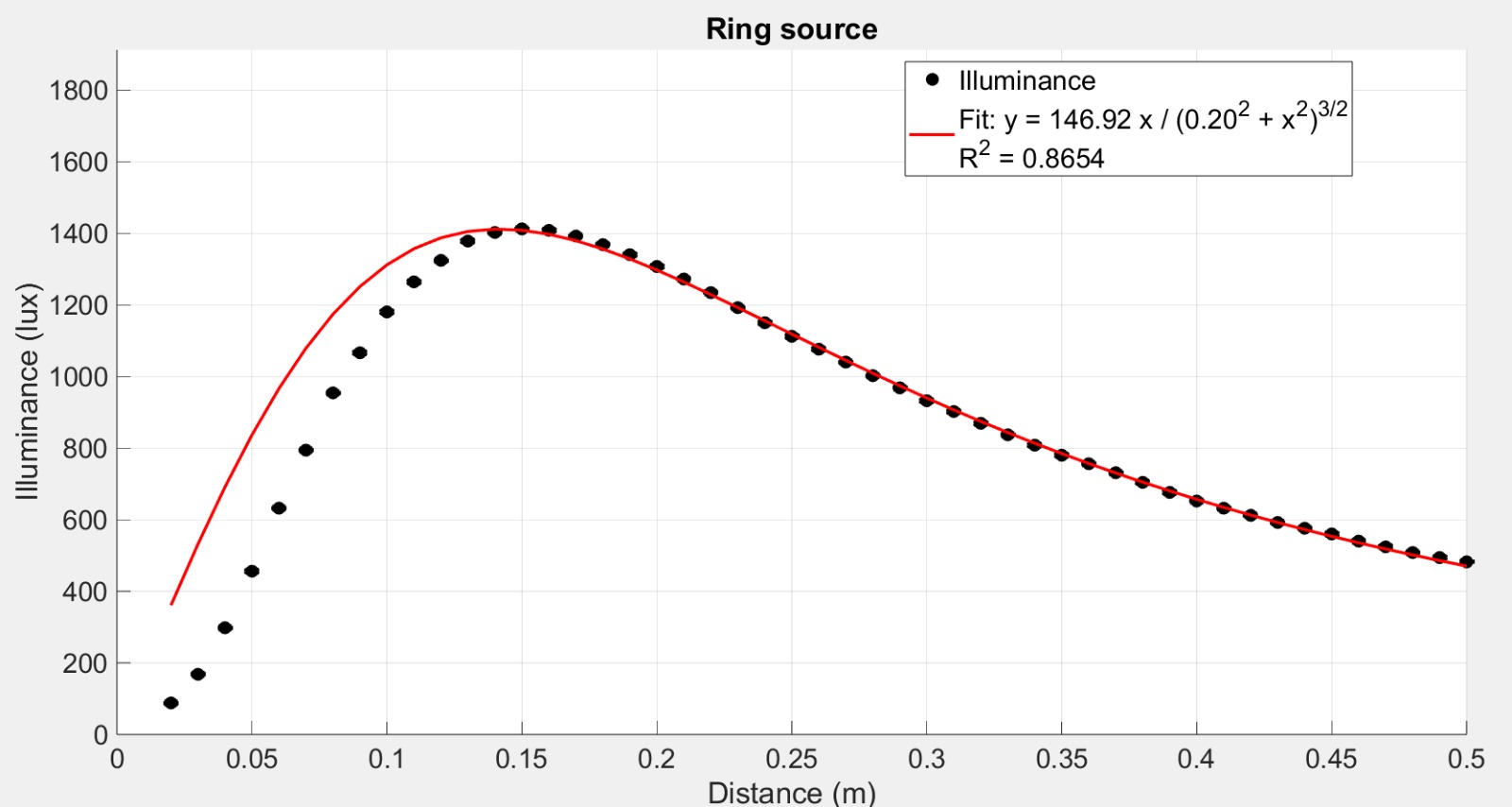}
\caption{
Illuminance as a function of distance for the ring light and the proposed non-linear fit, analogous to the one expected for the electric field of a charged ring. For distances less than 2R the measured illuminance is lower than that predicted by the model, due to the directivity of the sensor.
}
\label{fig:ringcurve}
\end{center}
\end{figure}

Finally, students analyzed a ring-shaped LED light source of radius R=13 cm, as illustrated in Fig.~\ref{fig:ringsetup}. In this case, the data were adjusted to a mathematical model analogous to the one expected for the electric field generated by a charged ring. As shown in Fig.~\ref{fig:ringcurve}, the model fits the data very well for distances greater than 2R. However, for distances less than 2R, the measured illuminances were lower than predicted by the model. This discrepancy allows students to understand an important factor such as the limitation imposed by the directivity of the light sensors. Specifically, the sensitivity of the sensors decreases as the angle of incidence deviates from the normal. When the sensor is positioned very close to the light ring, the incoming light arrives at relatively large angles, causing the sensor to record lower illuminances than it would if the light incidence were perpendicular.
Additionally, we note that ambient background light and reflections, clearly visible in Fig.~\ref{fig:ringsetup}, may also contribute to the discrepancies between the measured and predicted values. Although the experimental setup was designed to minimize such effects, one effective way to further reduce them is to place the light source as far as possible from nearby walls or tables, also ensuring that these surfaces are, whenever possible, covered with light-absorbing materials.

\section{Closing remarks}

In this activity, we proposed an exploration that integrates theoretical analysis with hands-on experimentation using accessible resources. Students  analyze a theoretical model and design a simple experimental setup using homemade materials, fostering creativity and problem-solving skills. By working with different source geometries, they  investigate the impact of the source's size, considering realistic scenarios where sources are neither infinite nor point-like. To measure illuminance, participants  utilize a smartphone—a readily available, easy-to-use device capable of providing accurate data—bridging everyday technology with scientific inquiry.
This activity was proposed to students as a distance activity, however, due to positive feedback from students, it was maintained even after the confinement measures were overcome. Although very simple in terms of materials and set up, it allows for great versatility and represents a valuable contribution to analyse and discuss, encouraging critical thinking and decision making both in the design of the experiment and discussion and analysis of the results.

Effective scientific learning can be fostered by engaging students with accessible and straightforward materials while studying theoretical models and acknowledging their limitations. Bridging the gap between theoretical and practical knowledge involves connecting these models with well-designed experiments, where setups and measurements are thoughtfully planned. Utilizing easily accessible and user-friendly measurement devices that deliver reliable results allows students to focus on the core principles without technical barriers. Encouraging cycles of experimentation, discussion, argumentation, and re-experimentation cultivates critical thinking and decision-making skills, empowering students to analyze outcomes critically and refine their approaches based on evidence and reasoning.

\section*{References}

\bibliography{mybib}

\end{document}